# EUROPIUM *s*-PROCESS SIGNATURE AT CLOSE-TO-SOLAR METALLICITY IN STARDUST SiC GRAINS FROM AGB STARS

Janaína N. Ávila [1,2,3], Trevor R. Ireland [1,2], Maria Lugaro [4], Frank Gyngard [5], Ernst Zinner [5], Sergio Cristallo [6], Peter Holden [1], Thomas Rauscher [7,8]

[1] Research School of Earth Sciences, The Australian National University, Canberra ACT 0200, Australia; janaina.avila@anu.edu.au

[2] Planetary Science Institute, The Australian National University, Canberra ACT 0200, Australia

[3] Astronomy Department/IAG, University of São Paulo, São Paulo SP 05508-090, Brazil

[4] Monash Centre for Astrophysics, Monash University, Clayton VIC 3800, Australia

[5] Laboratory for Space Sciences and the Department of Physics, Washington University, One Brookings Drive, St. Louis, MO 63130, USA

[6] Osservatorio Astronomico di Collurania, INAF, Teramo 64100, Italy

[7] Centre for Astrophysics Research, School of Physics, Astronomy, and Mathematics, University of Hertfordshire, Hatfield AL10 9AB, United Kingdom

[8] Department of Physics, University of Basel, 4056 Basel, Switzerland

Short Title:

EUROPIUM *s*-PROCESS SIGNATURE IN STARDUST SiC

ABSTRACT

Individual mainstream stardust silicon carbide (SiC) grains and a SiC-enriched bulk sample from the Murchison carbonaceous meteorite have been analyzed by the Sensitive High Resolution Ion Microprobe − Reverse Geometry (SHRIMP-RG) for Eu isotopes. The mainstream grains are believed to have condensed in the outflows of ~ 1.5 to 3 $M_\odot$ carbon-rich asymptotic giant branch (AGB) stars with close-to-solar metallicity. The $^{151}$Eu fractions [$fr(^{151}$Eu$) = ^{151}$Eu$/(^{151}$Eu$+^{153}$Eu$)$] derived from our measurements are compared with previous astronomical observations of carbon-enhanced metal-poor (CEMP) stars enriched in elements made by *slow* neutron captures (the *s*-process). Despite the difference in metallicity between the parent stars of the grains and the metal-poor stars, the $fr(^{151}$Eu$)$ values derived from our measurements agree well with $fr(^{151}$Eu$)$ values derived from astronomical observations. We have also compared the SiC data with theoretical predictions of the evolution of Eu isotopic ratios in the envelope of AGB stars. Because of the low Eu abundances in the SiC grains, the $fr(^{151}$Eu$)$ values derived from our measurements show large uncertainties, being in most cases larger than the difference between solar and predicted $fr(^{151}$Eu$)$ values. The SiC aggregate yields a $fr(^{151}$Eu$)$ value within the range observed in the single grains and provides a more precise result ($fr(^{151}$Eu$)$ = 0.54 ± 0.03, 95% conf.), but is approximately 12% higher than current *s*-process predictions. The AGB models can match the SiC data if we use an improved formalism to evaluate the contribution of excited nuclear states in the calculation of the $^{151}$Sm$(n, \gamma)$ stellar reaction rate.

Subject headings: dust, extinction — nuclear reactions, nucleosynthesis, abundances — stars: AGB and post-AGB — stars: carbon

1. INTRODUCTION

Measurements of the $^{151}$Eu isotope fraction [$fr(^{151}$Eu$) = {}^{151}$Eu$/(^{151}$Eu$+^{153}$Eu$)$] in stars are still scarce and only available for metal-poor stars (Sneden et al. 2002; Aoki et al. 2003a,b; Roederer et al. 2008), with the exception of the Solar System, where measurements on meteorites, as well as the Solar photosphere, give $fr(^{151}$Eu$) = 0.48$ (Lawler et al. 2001; Lodders 2003). Recent astronomical observations of the $^{151}$Eu isotope fraction in carbon-enhanced metal-poor (CEMP) stars enriched in the elements made by the *slow* neutron capture process (the *s*-process) have been shown to differ from the Solar System value (Aoki et al. 2003a), but to be in good agreement with *s*-process predictions in low mass asymptotic giant branch (AGB) stars of one-half solar metallicity (Arlandini et al. 1999). The values reported by Arlandini et al. (1999), however, were obtained using a theoretical $^{151}$Sm$(n, \gamma)^{152}$Sm cross section from Toukan et al. (1995) that is approximately 40% lower than recent experimental measurements (Abbondanno et al. 2004; Marrone et al. 2006; Wisshak et al. 2006). The higher $^{151}$Sm$(n, \gamma)^{152}$Sm cross section results in an enhanced neutron-capture channel on $^{151}$Sm and a lower production of $^{151}$Sm. Since $^{151}$Eu is the radiogenic product of the decay of $^{151}$Sm (with a half-life of 90 yrs), a higher $^{151}$Sm$(n, \gamma)$ cross section results in a lower predicted $fr(^{151}$Eu$)_{s\text{-process}}$ value, outside the range observed in CEMP stars.

Analyses of presolar stardust grains provide an opportunity to explore the Eu isotopic compositions in stars with a different range of metallicities than those of metal-poor stars. This is because the vast majority of stardust SiC grains (the "mainstream grains") are believed to have originated in the outflows of ~ 1.5 to 3 M$_\odot$ carbon-rich AGB stars with close-to-solar metallicity (Hoppe et al. 1994; Zinner et al. 2006). Here, we present the results of Eu isotopic analyses carried out with a Sensitive High Resolution Ion

Microprobe − Reverse Geometry (SHRIMP-RG). We analyzed thirteen large mainstream SiC grains (LS+LU fractions) and a SiC-enriched bulk sample (KJB fraction) extracted from the Murchison carbonaceous chondrite (Amari et al. 1994).

Previous Eu isotopic secondary ion mass spectrometry (SIMS) analyses by Terada et al. (2006) in two mainstream stardust SiC grains indicated $fr(^{151}Eu)$ values lower than $s$-process predictions and astronomical observations. The difference observed between the $fr(^{151}Eu)$ values obtained for SiC grains by Terada et al. (2006) and those obtained for CEMP stars by Aoki et al. (2003a) may indicate that $fr(^{151}Eu)$ depends on metallicity. However, the values reported by Terada et al. (2006) are even lower than the $fr(^{151}Eu)$ shown by the bulk Solar System. Since 98% of Eu in the Solar System is expected to have been produced by the *rapid* neutron capture nucleosynthesis ($r$-process), $fr(^{151}Eu)_\odot$ is commonly used as a proxy for $fr(^{151}Eu)_{r\text{-process}}$. Thus, it is surprising that the results presented by Terada et al. (2006) for grains believed to have condensed in the outflows of low mass carbon-rich AGB stars are closer to the $fr(^{151}Eu)$ expected for $r$-process nucleosynthesis than that expected for $s$-process nucleosynthesis. The results presented here will help to resolve these puzzles.

## 2. EXPERIMENTAL METHODS

### 2.1. Carbon, Nitrogen, and Silicon Isotopic Measurements

Carbon, N, and Si isotopic ratios for the KJB fraction have been previously reported by Amari et al. (2000), and are reproduced in Table 1. Two different mounts containing SiC grains from the LS+LU fractions were analyzed in the present study, namely LU and WU.

Grains from the LU mount have been analyzed previously with a modified Cameca IMS-3f ion microprobe at Washington University (St. Louis) for their C, N, and Si isotopic compositions as well as for their trace-element concentrations (Virag et al. 1992). Some of these grains were analyzed for their Ti, Ba, and W isotopic compositions with the Sensitive High Mass-Resolution Ion Microprobe at the Australian National University (Ireland et al. 1991; Ávila et al. 2012; 2013). Grains from the WU mount were analyzed for their C, N, and Si isotopic compositions with a Cameca NanoSIMS at Washington University. The NanoSIMS measurements were performed by rastering a ~ 100 nm $Cs^+$ primary beam (~ 1 pA) over a sample area of 2−10 $\mu m^2$ while secondary electrons and the negative secondary ions $^{12}C^-$, $^{13}C^-$, $^{28}Si^-$, $^{29}Si^-$, and $^{30}Si^-$ were simultaneously counted in electron multipliers. Subsequent to the C and Si isotopic measurements, nitrogen isotopes were measured as $CN^-$ ions at masses 26 ($^{12}C^{14}N^-$) and 27 ($^{12}C^{15}N^-$). Synthetic SiC was used for normalization of the C and Si isotopes, while a fine-grained mixture of SiC and $Si_3N_4$ was used for calibration of the N isotopes.

## 2.2. Europium Isotopic Measurements

Europium isotopic measurements in stardust SiC grains were carried out with the SHRIMP-RG at the Australian National University. We performed both "bulk analyses" on an aggregate of many grains from the KJB fraction and "single-grain analyses" on grains from the LS+LU fractions. Thirteen out of twenty five single grains investigated from the LS+LU fractions had sufficiently high Eu concentrations for isotopic analysis. SHRIMP-RG measurements were performed with an $O_2^-$ primary beam of 2–5 nA focused to sputter an area of ~ 20 μm in diameter. Secondary ions were extracted at 10 keV and measured by single collector analysis on the ETP$^{TM}$ multiplier in magnetic peak-jumping mode. The acquisition time for each grain was ~ 5 min, which consisted of 5–6 scans through the following peaks: $^{138}$Ba$^+$, $^{139}$La$^+$, $^{140}$Ce$^+$, $^{151}$Eu$^+$, and $^{153}$Eu$^+$. We systematically bracketed three unknowns by a suite of standard reference materials (NIST-610 silicate glass, USGS BCR-2G silicate glass, and a SiC ceramic doped with heavy elements, Ávila et al. 2013).

The measurement of $^{151}$Eu$^+$ and $^{153}$Eu$^+$ in stardust SiC grains is challenging because of potential isobaric interferences. The secondary ion signals at masses 151 and 153 consist of contributions from $^{151}$Eu$^+$ and $^{153}$Eu$^+$, as well as from BaO$^+$ ($^{135}$Ba$^{16}$O$^+$ and $^{137}$Ba$^{16}$O$^+$, respectively). In order to resolve the $^{151}$Eu$^+$ and $^{153}$Eu$^+$ peaks from monoxide interferences, SHRIMP-RG was operated at a mass resolving power of m/Δm= 8000 (at 10% peak height). At this level, the BaO$^+$ species were well resolved from the Eu isotopes, with insignificant tailing contributions.

Further investigation of a "pure" synthetic SiC revealed the presence of molecular interferences in the mass region of interest. The ratio 151/153 in the "pure" synthetic SiC was found to be ~ 0.5 ($fr$($^{151}$Eu) = 0.33), which is considerably different from the Solar System $^{151}$Eu/$^{153}$Eu ratio of 0.916 (Lodders 2003). The mass offset between the atomic species ($^{151}$Eu$^+$ and $^{153}$Eu$^+$) and the molecular interferences is too small to be resolved by mass separation without drastically compromising the secondary ion yields. Therefore, to suppress molecular ion contributions to the atomic species, we used an energy filtering technique similar to the one described in Ávila et al. (2013). We found that for an energy offset of approximately 21 eV, selected based on the momentum spectrum at mass region 151 a.m.u. of ions sputtered from the NIST-610 silicate glass and a "pure" synthetic SiC, the molecular interferences were completely excluded from collection; however, the intensity of the secondary ion signal dropped by approximately one order of magnitude. A combination of high mass resolution and energy filtering was used in all analyses presented in this work. The NIST-610 silicate glass and a SiC ceramic doped with heavy elements (Ávila et al. 2013) were used to correct for instrumental mass fractionation.

3. RESULTS

The LS+LU SiC grains studied here show Si, C, and N isotopic compositions in the range displayed by mainstream grains (Table 1) suggesting that these grains formed in the outflows of low mass C-rich AGB stars with close-to-solar metallicity. The $^{151}$Eu/$^{153}$Eu ratios measured in single SiC grains from the LS+LU fractions range from 0.95 to 1.62 (Table 1), compared to the Solar System $^{151}$Eu/$^{153}$Eu of 0.916 (Lodders 2003). The $fr(^{151}$Eu$)$ values derived from our measurements range from 0.49 to 0.61 (Fig. 1 and Table 1). The weighted mean $^{151}$Eu/$^{153}$Eu of the SiC-enriched bulk sample (KJB fraction), based on 4 measurements, is $1.20 \pm 0.14$ (95% conf.), which corresponds to a $fr(^{151}$Eu$)$ value of $0.54 \pm 0.03$ (95% conf.). All single SiC grains show $fr(^{151}$Eu$)$ values higher than the Solar System value, however, only one out of thirteen grains shows a $fr(^{151}$Eu$)$ value that differs from the Solar System value by more than 2σ.

Our results are clearly at odds with previous SIMS determinations in two mainstream stardust SiC grains from low-mass AGB stars that indicated lower-than-solar $fr(^{151}$Eu$)$ values (Terada et al. 2006). The $fr(^{151}$Eu$)$ values derived by Terada et al. (2006) for the two SiC grains are ~ 0.43. We suspect that the inconsistency between our measurements and those of Terada et al. (2006) may be due to differences in the experimental approach used. We have shown previously that energy filtering, i.e., the selection of an appropriate energy window for the secondary ions on the basis of the momentum spectrum, is essential for suppressing unwanted interferences in the mass region of the Eu isotopes in a SiC matrix. The data of Terada et al. (2006) were obtained by using high-mass resolution alone, which we have found to be insufficient to remove molecular interferences that can significantly affect the Eu isotope measurements. Given the low value measured by Terada et al. (2006) and the similarly low ratio we have measured in the pure SiC standard, we suspect that their measurements are compromised.

Since the SiC-enriched bulk sample (KJB fraction) yields a $fr(^{151}Eu)$ value that is within the range observed in the single grains, but considerably more precise, we will focus the following discussion on this result.

## 4. DISCUSSION

In AGB stars, the *s*-process nucleosynthesis is responsible for the production of elements heavier than Fe. The main neutron source for the *s*-process in the He-intershell of AGB stars is the $^{13}C(\alpha, n)^{16}O$ reaction which operates under radiative conditions at relatively low temperatures (T ~ $0.9 \times 10^8$ K) during the intervals between thermal pulses, and results in low neutron densities (~ $10^6 - 10^7$ neutrons cm$^{-3}$). During thermal pulses in low-mass AGB stars, the $^{22}Ne(\alpha, n)^{25}Mg$ reaction is marginally activated when the maximum temperature at the bottom of the He-burning shell reaches T ~ $3 \times 10^8$ K, producing a small neutron burst with a high neutron density peak (up to ~ $10^{10}$ neutrons cm$^{-3}$).

Branching points on the *s*-process path at $^{151}Sm$, $^{153}Sm$, $^{152}Eu$, and $^{153}Gd$ affect the production of $^{151}Eu$ and $^{153}Eu$ isotopes in the He-intershell of AGB stars (Fig. 2a). The competition between neutron capture and *β*-decay at these branching points can be expressed by a branching factor ($f_n$), calculated from:

$$f_n = \frac{\lambda_n}{(\lambda_n + \lambda_\beta)}$$

where $\lambda_n = N_n\, v_T\, \langle\sigma\rangle$ and $\lambda_\beta = \ln2/t_{1/2}$ are the neutron capture rate and the *β*-decay rate, respectively. Here, $N_n$, $v_T$, $\langle\sigma\rangle$, and $t_{1/2}$ are the neutron density, the thermal velocity, the Maxwellian averaged (n, γ) cross section (MACS), and the half-life, respectively. Branching factors were calculated for $^{151}$Sm, $^{153}$Sm, $^{152}$Eu, and $^{153}$Gd as a function of neutron density for thermal energies of kT = 8 keV (T ~ 0.9 × 10$^8$ K) and kT = 23 keV (T ~ 2.7 × 10$^8$ K). The results are given in Fig. 2b and c. We used the *β*-decay rates reported by Takahashi & Yokoi (1987) and the latest accepted neutron-capture rates from Dillmann et al. (2006). Note that the branching at $^{152}$Eu is peculiar, since it involves the competition between neutron capture, *β*-decay, and electron capture (EC). The branching at $^{153}$Gd is governed by electron capture and neutron capture.

During the interpulse periods most of the *s*-process flow proceeds via the sequence $^{151}$Sm(n, γ) $^{152}$Sm(n, γ) $^{153}$Sm(*β*, *ν*) $^{153}$Eu bypassing $^{151}$Eu. After the neutron flux is extinguished, $^{151}$Eu is fed by the radioactive decay of $^{151}$Sm. During the thermal pulses, the *β*-decay rate of $^{151}$Sm is increased by a factor of ~ 26 (Takahashi & Yokoi 1987) and the *s*-process path can branch toward $^{151}$Eu. However, the neutron-capture channel on $^{151}$Sm is also enhanced since higher neutron densities are produced together with higher temperatures. The branching points at $^{153}$Gd and $^{153}$Sm are largely open at these high neutron densities, feeding $^{154}$Gd and $^{154}$Sm, respectively, which result in a smaller *s*-process contribution to $^{153}$Eu. The AGB model predictions shown in Fig. 3 demonstrate that the final result is a very mild enhancement of $fr(^{151}$Eu) values in the stellar envelope.

In Fig. 3 the stardust SiC-enriched bulk data are compared to the *s*-process AGB predictions from the FRUITY database (for details see Cristallo et al. 2009, 2011). In all stellar models, the $fr(^{151}Eu)_{envelope}$ is not significantly different from the solar value. The close-to-solar metallicity models ($Z = 0.01$ and $Z = 0.014$, Figs. 3a and 3b, respectively) show $fr(^{151}Eu)_{envelope}$ approximately 2% higher than the $fr(^{151}Eu)_{\odot}$ value (Lodders 2003). The low-metallicity models ($Z = 0.0001$ and $Z = 0.0003$, Figs. 3c and 3d, respectively), on the other hand, show $fr(^{151}Eu)_{envelope}$ values that are about 4% lower than the $fr(^{151}Eu)_{\odot}$ value. It is important to note that Cristallo et al. (2009, 2011) adopted the experimental $^{151}Sm(n, \gamma)$ cross section reported by Marrone et al. (2006), which is significantly higher than the value used by Arlandini et al. (1999). This higher cross section results in a lower $fr(^{151}Eu)$ as more $^{151}Sm$ is destroyed by neutron captures. The $fr(^{151}Eu)$ derived for the SiC-enriched bulk sample (KJB) shows a higher $fr(^{151}Eu)$ value than those predicted by Cristallo et al. (2009, 2011).

The $fr(^{151}Eu)$ values derived from our measurements agree well with $fr(^{151}Eu)$ values derived from astronomical observations (Fig. 1). Aoki et al. (2003a) reported $fr(^{151}Eu)$ for two CEMP stars, LP 625-44 and CS 31062-050, which show excesses of *s*-process elements. These stars are very metal-poor subgiants with [Fe/H] of −2.7 (LP 625-44, Aoki et al. 2000) and −2.4 (CS 31062-050, Aoki et al. 2002). Both stars show variations of their radial velocities, indicating that they belong to binary systems (Aoki et al. 2000, 2003a). According to the classification suggested by Jonsell et al. (2006), LP 625-44 and CS 31062-050 are CEMP-*r*+*s* stars, showing both *s*- and *r*-process enhancements. One possible scenario to explain the enrichment observed in CEMP-*r*+*s* stars is to assume that the parent cloud of the binary system was already enriched in *r*-process elements, while the *s*-process elements (and also carbon) were the result of mass transfer from an AGB star to the lower-mass companion (e.g., Bisterzo et al. 2010). Lugaro et al. (2012) instead propose that the composition of CEMP-$r$+$s$ stars is the result of a neutron-capture process in-between the $s$-process and the $r$-process. The $fr(^{151}Eu)$ values derived by Aoki et al. (2003a) for LP 625-44 and CS 31062-050 are 0.60 and 0.55, respectively. These values are clearly higher than $fr(^{151}Eu)_\odot$ of 0.48 and current AGB model predictions from Cristallo et al. (2009, 2011).

To address this mismatch we performed several tests varying the neutron-capture cross sections and beta-decay rates that affect $fr(^{151}Eu)$ and found that the $^{151}Sm(n, \gamma)$ reaction rate plays the major role in setting the predicted $fr(^{151}Eu)$ value. In Fig. 4 we show $fr(^{151}Eu)$ in the envelope of a 2 $M_\odot$ model with Z = 0.01 computed with different theoretical $^{151}Sm(n, \gamma)$ reaction rates and their uncertainties. Models have been calculated with the FUll-Network Stellar Evolutionary Code (Straniero et al. 2006 and references therein). The results labeled as "M06" were obtained using the experimental data by Marrone et al. (2006), multiplied by the theoretical stellar enhancement factor (SEF) from Bao et al. (2000). Marrone et al. (2006) reported an experimental uncertainty on the order of 5%, which results in a narrow predicted range of $fr(^{151}Eu)$. In previous investigations, applying the SEF was thought to account for the contribution of excited nuclear states in stellar environments. It has been shown that this is not the case by Rauscher et al. (2011), and Rauscher (2012) presented an improved approach, also accounting for the uncertainties in the stellar rates. This leads to error bars larger than those given by the experiments due to the remaining theoretical uncertainties in the excited state contributions to the stellar rate. Using this approach (labeled "R12" in Fig. 4), we obtained a larger allowed range of $fr(^{151}Eu)$, which easily covers the values observed both in CEMP stars and in SiC grains.

## 5. CONCLUSION

We presented new Eu isotopic data obtained on stardust SiC grains with SHRIMP-RG and compared them to previous astronomical observations in CEMP stars enriched in *s*-process elements and with *s*-process AGB model predictions. Despite the difference in metallicity between the parent stars of the grains and the metal-poor stars, we found good agreement between the grains and the stellar data. Because of large uncertainties in the single grain data, the result of low Eu concentrations, only in one case is the $fr(^{151}\text{Eu})$ value derived for a single grain higher than $fr(^{151}\text{Eu})_\odot$ by more than 2σ. The $fr(^{151}\text{Eu})$ value derived for the KJB aggregate is, on the other hand, very well constrained, and matches those of CEMP stars, but is approximately 12% higher than current *s*-process predictions. Our new data can only be matched when the uncertainties predicted with the model of Rauscher (2012) are applied and support this more accurate approach of computing neutron-capture cross sections in stellar environments.

## ACKNOWLEDGEMENTS

J.N. Ávila acknowledges support by CNPq grants #200081/2005-5 and #150570/2011-2. T. R. Ireland acknowledges support by ARC grants DP0342772 and DP0666751. M. Lugaro acknowledges the support of the ARC via a Future Fellowship and of Monash University via a Monash fellowship. E. Zinner acknowledges support by NASA grant NNX11AH14G. S. Cristallo acknowledges financial support from the FIRB2008 program (RBFR08549F-002) and from the PRIN-INAF 2011 grant. T. Rauscher acknowledges support by the Swiss NSF, the EUROCORES EuroGENESIS research program, and the ENSAR/THEXO European FP7 program. We thank an anonymous referee for their comments and Fred Rasio for handling of this manuscript.

FIGURE CAPTIONS

Figure 1: $fr(^{151}Eu)$ values measured for single grains (LS+LU fractions) and the SiC-enriched bulk sample (KJB fraction). The $fr(^{151}Eu)$ values are compared with two CEMP-*r*+*s* stars, LP 625-44 and CS 31062-050 (Aoki et al., 2003a). The black dashed line indicates $fr(^{151}Eu)_\odot$ given by Lodders (2003). The grey band corresponds to the $fr(^{151}Eu)$ value (weighted mean ± 1σ) derived from measurements of the SiC-enriched bulk sample (KJB fraction). Error bars are 1σ.

Figure 2: (a) Part of the nuclide chart showing the *s*-process nucleosynthesis path in the region of Sm-Eu-Gd. Percent abundances in the Solar System (non-italic) are shown for each stable isotope (solid boxes) and laboratory half-lives (italic) for each unstable isotope (dashed line boxes). The main *s*-process path is shown as a bold line and branches and secondary paths are shown as finer lines. The *s*-only isotopes $^{150}$Sm, $^{152}$Gd, and $^{154}$Gd are indicated by bold boxes. (b, c) Branching factors ($f_n$) at $^{151}$Sm, $^{153}$Sm, $^{152}$Eu, and $^{153}$Gd are shown as a function of neutron density at temperatures kT = 8 keV and 23 keV. All values are calculated for an electron density of $5 \times 10^{26}$ cm$^{-3}$. The branching factor (given in %) indicates the probability that the unstable isotope captures a neutron rather than decays. The grey areas in (b) and (c) correspond to the conditions typically found during interpulse and thermal pulse phases, respectively, in low-mass AGB stars.

Figure 3: Evolution of $fr(^{151}Eu)_{envelope}$ as a function of the thermal pulse number for theoretical models of AGB stars of masses $M$ = 1.3, 1.5, 2, 2.5, and 3 M$_\odot$ and metallicities $Z$ = 0.0001, 0.0003, 0.01, and 0.014 (Cristallo et al., 2009, 2011). Symbols are only shown for C/O > 1 in the stellar envelope. The grey band in (a)

and (b) corresponds to the $fr(^{151}Eu)$ value (weighted mean ± 1σ) derived from measurements of the SiC-enriched bulk sample (KJB fraction) and the grey band in (c) and (d) corresponds to the $fr(^{151}Eu)$ value (weighted mean ± 1σ) derived from observations of CEMP-$r$+$s$ stars (LP 625-44 and CS 31062-050; Aoki et al., 2003a).

Figure 4: Evolution of $fr(^{151}Eu)$ in the envelope as a function of $\Delta t_{TP-AGB}$ (time from the beginning of the TP-AGB phase) for a $M = 2$ M$_\odot$ and $Z = 0.01$ AGB model using two different $^{151}Sm(n, \gamma)$ reaction rates. The shaded area (plotted only for the C-rich phases of the evolution) represents the associated uncertainties. M06 represents the experimental rate from Marrone et al. (2006) multiplied by the SEF from Bao et al. (2000). R12 represents the stellar rate computed using the method of Rauscher (2012) including the experimental data from Marrone et al. (2006).

Table 1: C, N, Si, and Eu isotopic compositions of stardust SiC grains from the KJB and LS+LU fractions. Errors are 1σ.

| Spot/ Grain | Size (μm) | $^{12}C/^{13}C$ ± 1σ | $^{14}N/^{15}N$ ± 1σ | $\delta^{29}Si/^{28}Si$ [a] ± 1σ (‰) | $\delta^{30}Si/^{28}Si$ [a] ± 1σ (‰) | $^{151}Eu/^{153}Eu$ ± 1σ | $fr\,(^{151}Eu)$ [b] ± 1σ | Eu ppm |
|---|---|---|---|---|---|---|---|---|
| Solar | | 89.0 | 459 | 0 | 0 | 0.92 | 0.48 | |
| Murchison SiC-enriched bulk sample (KJB fraction) | | | | | | | | |
| KJB [c] | 0.49 | 37.0 ± 0.4 | 521 ± 60 | 24.6 ± 1.3 | 37.8 ± 3.4 | | | |
| KJB-01 | | | | | | 1.19 ± 0.22 | 0.54 ± 0.04 | 0.192 |
| KJB-02 | | | | | | 1.20 ± 0.11 | 0.55 ± 0.03 | 0.198 |
| KJB-03 | | | | | | 1.21 ± 0.14 | 0.55 ± 0.03 | 0.216 |
| KJB-04 | | | | | | 1.18 ± 0.12 | 0.54 ± 0.03 | 0.143 |
| Weighted averaged | | | | | | 1.20 ± 0.07 | 0.54 ± 0.02 | |
| Murchison single SiC grains (LS+LU fractions) | | | | | | | | |
| **Mount WU** | | | | | | | | |
| WU-01 | 8 x 9 | 57.5 ± 0.4 | 434 ± 24 | 98.6 ± 5.4 | 73.5 ± 9.7 | 1.02 ± 0.15 | 0.51 ± 0.04 | 0.084 |
| WU-03 | 10 x 12 | 48.3 ± 0.3 | 461 ± 37 | 38.4 ± 4.7 | 51.4 ± 9.1 | 0.95 ± 0.05 | 0.49 ± 0.01 | 0.006 |
| WU-04 | 26 x 34 | 49.3 ± 0.4 | 304 ± 14 | 32.4 ± 4.8 | 29.9 ± 9.1 | 1.28 ± 0.36 | 0.56 ± 0.08 | 0.002 |
| WU-05 | 6 x 8 | 84.2 ± 0.6 | n.a. | 9.0 ± 4.6 | 31.9 ± 9.0 | 1.17 ± 0.19 | 0.54 ± 0.04 | 0.051 |
| WU-14 | 4 x 4 | 81.8 ± 0.6 | 377 ± 23 | 37.0 ± 4.8 | 50.8 ± 9.2 | 0.97 ± 0.09 | 0.49 ± 0.02 | 0.111 |
| WU-16 | 7 x 9 | 65.4 ± 0.5 | 416 ± 14 | 35.5 ± 4.8 | 25.7 ± 9.0 | 1.04 ± 0.12 | 0.51 ± 0.03 | 0.046 |
| WU-17 | 6 x 6 | 52.2 ± 0.4 | 413 ± 23 | 110.3 ± 5.0 | 101.2 ± 9.6 | 1.62 ± 0.40 | 0.62 ± 0.07 | 0.088 |
| WU-24 | 7 x 9 | 90.4 ± 0.7 | 340 ± 30 | −15.7 ± 4.5 | 0.4 ± 8.7 | 1.18 ± 0.25 | 0.54 ± 0.05 | 0.016 |
| WU-54 | 6 x 6 | 51.6 ± 0.4 | 345 ± 19 | 69.3 ± 5.4 | 64.2 ± 9.7 | 1.10 ± 0.39 | 0.52 ± 0.05 | 0.044 |
| **Mount LU** [d] | | | | | | | | |
| LU-29b | 16 x 20 | 49.3 ± 0.5 | 610 ± 17 | 34.8 ± 3.1 | 38.2 ± 3.5 | 1.54 ± 0.52 | 0.61 ± 0.06 | 0.006 |
| LU-30a | 8 x 15 | 48.8 ± 0.4 | 467 ± 17 | 42.4 ± 2.5 | 44.5 ± 3.2 | 1.46 ± 0.47 | 0.59 ± 0.06 | 0.010 |
| LU-32 | 5 x 13 | 63.0 ± 0.4 | 1088 ± 14 | 55.3 ± 2.5 | 47.8 ± 3.2 | 1.54 ± 0.46 | 0.61 ± 0.10 | 0.012 |
| LU-33 | 7 x 15 | 48.5 ± 0.3 | 1314 ± 21 | 39.5 ± 2.5 | 43.0 ± 3.1 | 1.23 ± 0.32 | 0.55 ± 0.05 | 0.024 |

[a] $\delta^{i}Si/^{28}Si\,(‰) = [(^{i}Si/^{28}Si)_{measured}/(^{i}Si/^{28}Si)_{solar} -1] \times 10^3$.
[b] $fr(^{151}Eu) = {}^{151}Eu/(^{151}Eu+{}^{153}Eu)$.
[c] C, N, and Si isotopic data of grains from KJB fraction reproduced from Amari et al. (2000).
[d] C, N, and Si isotopic data of grains from mount LU reproduced from Virag et al. (1992).
n.a. = not analyzed.

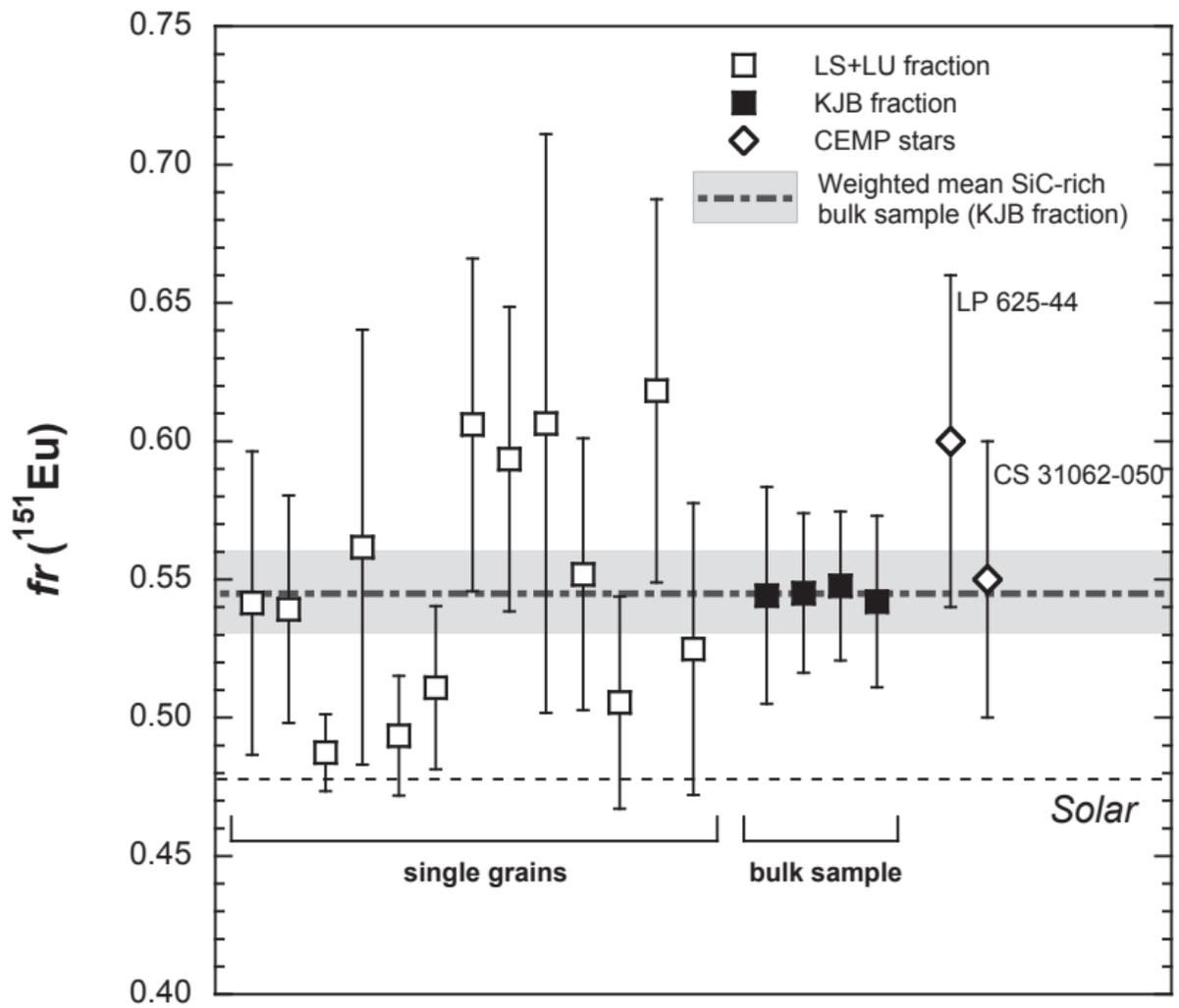

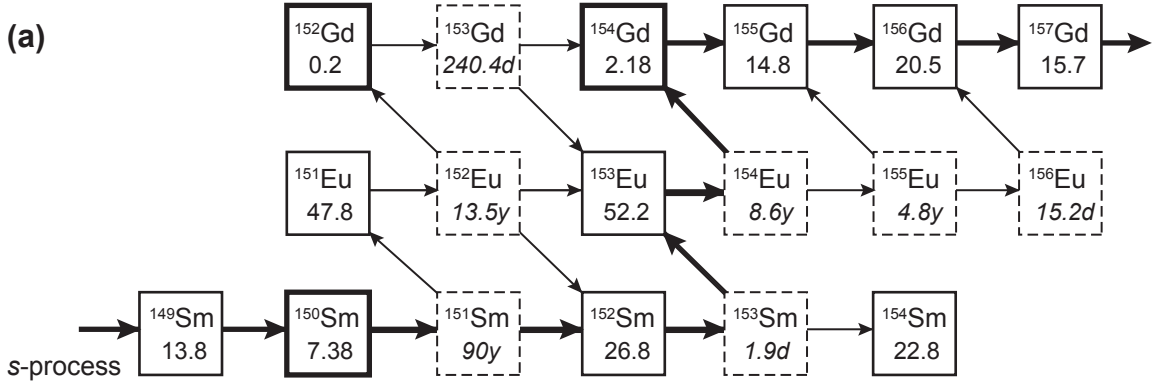
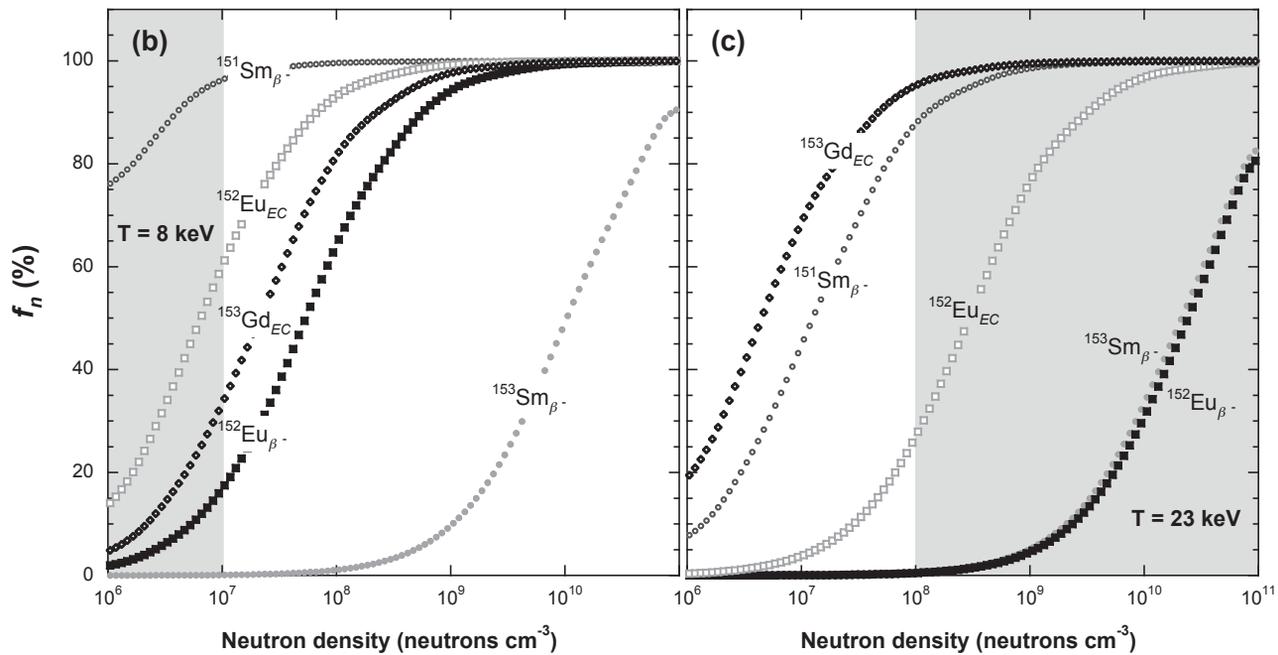

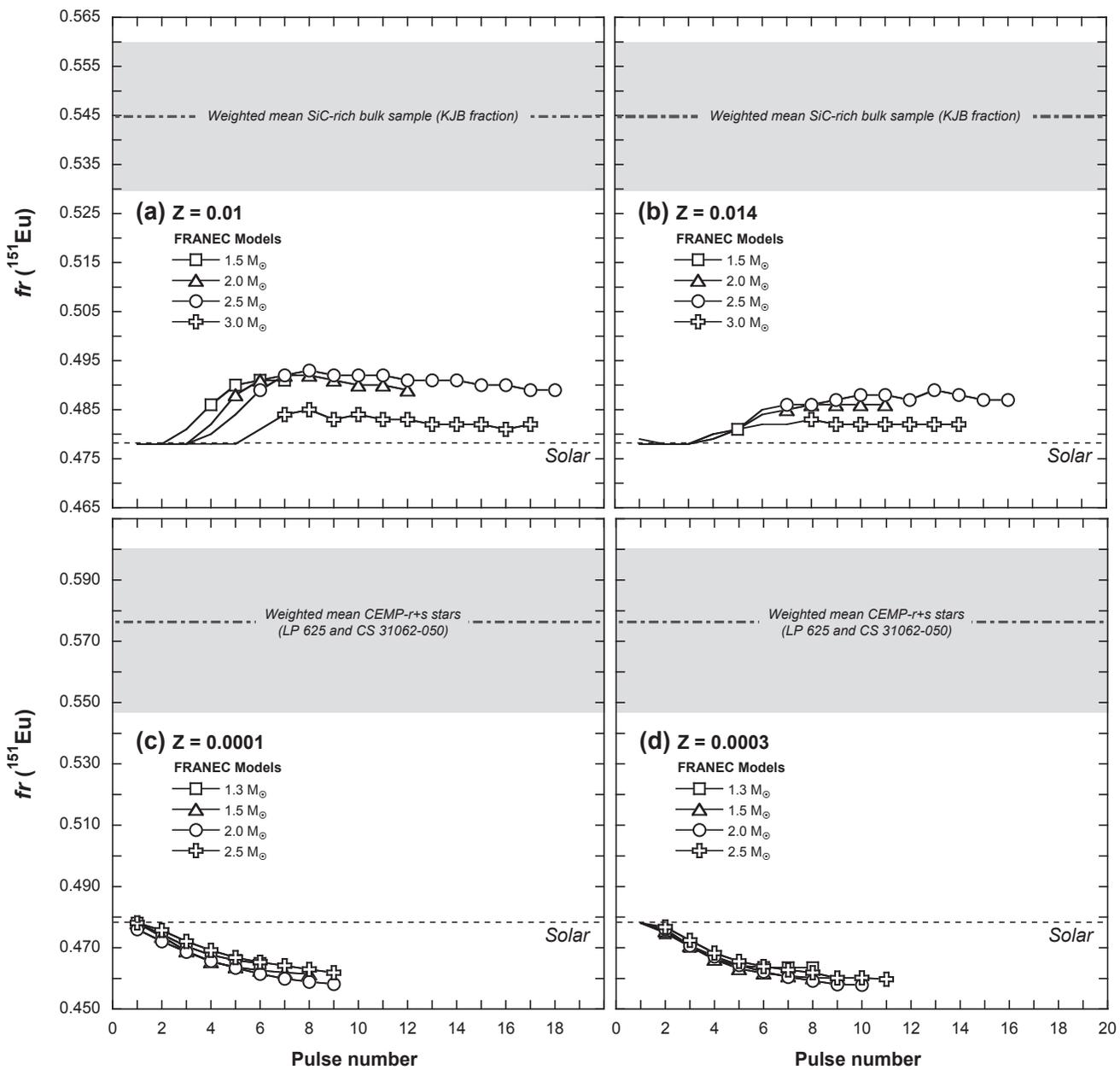

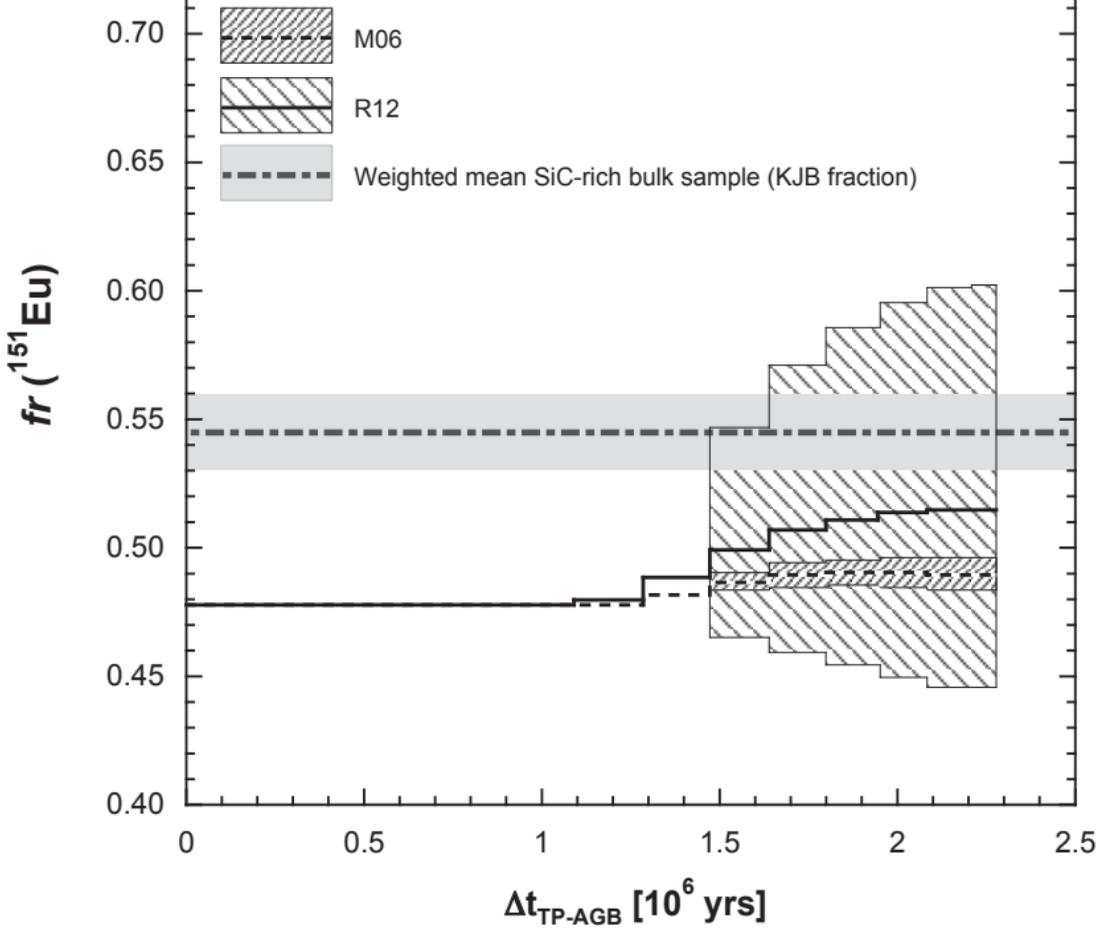